# The Impact of Social Isolation on Subjective Cognitive Decline in Older Adults: A Study Based on Network Analysis and Longitudinal Model

Yingchen Liu [a*], Haixin Jiang [a*]

[a] Faculty of Psychology, Southwest University, Chongqing 400715, China

[*] Corresponding author: Haixin Jiang, Faculty of Psychology, Southwest University, Chongqing 400715, China, e-mail address: iris9460@email.swu.edu.cn

**Research Transparency Statement**

**Conflicts of interest:** All authors declare no conflicts of interest.

**Acknowledgments:** This work was supported by the National Natural Science Foundation of China (31971007, 32371109, 71942004).

**Ethics statement:** This study was reviewed and approved by the Ethics Committee of the Faculty of Psychology, Southwest University (No. H23159, 2023) following the Declaration of Helsinki and was pre-registered in As Predicted (https://aspredicted.org/bk8e5.pdf). All participants were fully informed of the purpose of the study and completed the questionnaire after informed consent.

**Abstract:**

Social isolation (SI) in older adults has emerged as a critical mental health concern, with established links to cognitive decline. While depression is hypothesized to mediate this relationship, the longitudinal mechanisms remain unclear. This study employed network analysis and cross-lagged modeling to examine these relationships during pandemic-related social restrictions. We collected self-reported data from 1,230 older adults ($M_{age}$ = 64.49, $SD$ = 3.84) assessing their experiences across three temporal periods (during lockdown, immediately post-lockdown, and 6 months later). Network analysis identified depression (particularly PHQ-9 item 9) as central nodes bridging SI and subjective cognitive decline (SCD). Longitudinal analyses revealed: 1) T1 SI predicted T2 depression; 2) T2 depression predicted T3 SCD. These patterns held for both online and offline SI, with good model fit. Our findings demonstrate depression's mediating role in the "SI-depression-SCD" pathway, highlighting how reduced social connections may gradually impair cognitive self-assessment through depression. The results underscore the universal mental health impact of SI, regardless of interaction modality (online/in-person). These insights suggest that combined interventions targeting both social connection and depression management may be particularly effective for mitigating age-related cognitive decline.

**Introduction**

With the continuous progression of global population aging, mental health issues among older adults have become increasingly prominent, emerging as a critical research focus in psychology and public health. Extensive epidemiological data indicate a high prevalence of social isolation among older adults (studies reporting rates ranging from 10% to 43%; Smith & Hirdes, 2009). For instance, approximately 25% of community-dwelling older adults in the United States experience varying degrees of social isolation, with 43% reporting significant feelings of loneliness (National Academies of Sciences, Engineering, and Medicine, 2020). Existing research has demonstrated the robust protective effects of active social engagement on older adults' physical and mental health. For example, maintaining strong social relationships can increase life expectancy by up to 50% (Holt et al., 2010). A 70-year longitudinal study further confirmed that the quality of social relationships predicts health outcomes more strongly than do many traditional biological markers or socioeconomic factors (Vaillant et al., 2008). Conversely, social isolation in older adults is closely associated with numerous adverse psychological consequences, including depression, anxiety, and cognitive decline, and may even significantly elevate the risk of Alzheimer's disease (Nicholson, 2012).

First, substantial empirical evidence suggests a significant association between social isolation and cognitive decline. Individuals with smaller social networks often exhibit more pronounced cognitive impairment (Röhr et al., 2020). During the COVID-19 pandemic revealed that both subjectively perceived loneliness and objectively measured lack of social interaction impair affective and social cognitive functioning to varying degrees(Bland et al., 2021). Dynamic changes in social networks are closely linked to alterations in cognitive function: as the frequency of social interactions gradually decreases and social network size continues to shrink, individuals tend to experience progressive cognitive decline (Röhr et al., 2020). A significant positive association between subjectively perceived social isolation and long-term cognitive deterioration was found through a longitudinal study, a finding that has been validated in multiple studies in different cultural contexts (Hajek et al., 2020). Further, social isolation negatively affects cognition across multiple cognitive domains (Evans et al., 2018), including memory, executive function, and information processing speed.. The COVID-19 pandemic provided compelling natural experimental evidence, as both perceived loneliness and objective interaction deficits were found to differentially impair socio-cognitive functioning (Bland et al., 2021), while coercive measures exacerbated pre-existing isolation (Steijvers et al., 2022). Crucially, the post-restriction period offered unique insights into how social network dynamics mechanistically influence mental health trajectories in aging populations.

Subjective Cognitive Decline (SCD) is a distinctive cognitive state prevalent among older adults, characterized by persistent self-perceived cognitive deterioration despite the absence of objective neuropsychological abnormalities (Jessen et al., 2014; Molinuevo et al., 2017). Although SCD does not meet clinical diagnostic criteria for mild cognitive impairment or dementia, its subjective experience can reflect early trends of cognitive change in older adults, offering unique predictive value for cognitive decline trajectories. Clinical epidemiological studies suggest that SCD may serve as an early warning sign of Alzheimer's disease, with affected individuals exhibiting significantly elevated relative risk of progression to dementia (Mitchell et al., 2014). Accurate identification of older adults with SCD provides a critical time window for early intervention (Smart et al., 2016), underscoring its clinical screening significance. Thus, this study selects SCD as a key indicator of cognitive status to examine its relationship with social isolation.

Second, depression, a vital metric for assessing mental health in older adults, plays a pivotal role in cognitive decline. Age-related reductions in social support may trigger depression, forming a critical pathway influencing cognitive function (Gallagher et al., 2018). Substantial clinical evidence indicates that depression not only markedly diminishes quality of life across multiple domains but may also establish a mutually reinforcing vicious cycle with cognitive dysfunction. Existing research reveals the depression-cognition link through both static and dynamic dimensions: Cross-sectional studies demonstrate significant correlations between depression and cognitive impairment (Chan et al., 2020), while longitudinal designs further confirm that depressive states prospectively predict subsequent cognitive decline (Hu et al., 2020). Both depression and cognitive function in older adults exhibit temporal fluctuations, with significant cross-lagged predictive effects between the two(Barrenetxea et al., 2022). This association may stem from loneliness induced by deficient social relationships(Hawkley et al., 2008). The experience of large-scale pandemic lockdowns likely exacerbated the vicious cycle between social isolation and depression among older adults, with lingering effects long after restrictions were lifted. Consequently, dynamically modeling the relationships among social isolation, depression, and SCD is essential (Jiang et al., 2025).

Building on this evidence, this study employs a hybrid cross-sectional and longitudinal design to incorporate depression into the relationship model between social isolation and SCD, specifically investigating their potential mediating role. The research aims to elucidate how their relationships evolve during lockdowns, immediately after restrictions eased, and six months post-lockdown. Findings may inform targeted interventions to mitigate cognitive decline in older adults by systematically enhancing social support to alleviate depression. The hypotheses are as follows:

Hypothesis 1 (H1): Significant correlations exist among social isolation, depression, and subjective cognitive decline in older adults.

Hypothesis 2 (H2): depression mediate the relationship between social isolation and subjective cognitive decline, wherein diminished social support exacerbates depression, thereby driving subjective cognitive decline.

**Method**

This study recruited a total of 1,230 older adults from across China through the online "Wenjuanxing" platform, comprising 663 males (53.90%) and 567 females (46.10%). The mean age of participants was 64.49 years (SD = 3.84). The inclusion criteria for older adult participants were: 1) age ≥ 60 years; 2) community-dwelling older adults; 3) a total score of ≥ 24 on the Mini-Mental State Examination (MMSE; Folstein et al.,

1975). Table 1 presents the demographic characteristics of the participants.

Table 1: Basic demographic information

| Basic demographic information | *Mean*(SD or ratio%) |
|---|---:|
| Number of subjects | 1230 |
| Age | 64.49(3.84) |
| Gender | |
|   Male | 663(53.90%) |
|   Female | 567(46.10%) |
| Level of education (years) | 11.83(4.19) |
| Household income (per capita per month/RMB) | |
|   < 1000 | 10(0.81%) |
|   1000-3000 | 73(5.93%) |
|   3001-5000 | 301(24.47%) |
|   5001-10000 | 498(40.49%) |
|   > 10000 | 348(28.29%) |
| Substance use (sleep medications) | |
|   Never | 1031(83.82%) |
|   Less than 1 time/week | 114(9.27%) |
|   1-2 times/week | 76(6.18%) |
|   ≥ 3 times/week | 9(0.73)% |

We conducted an initial online survey to collect participants' basic demographic information, including name, gender, residence location, age, educational level, and general health status. Additionally, we assessed levels of social isolation, depression, and subjective cognitive decline.

Social Isolation: The Lubben Social Network Scale-6 (LSNS-6; Lubben et al., 2006) was used to evaluate participants' social support networks over the past month. This scale objectively measured social isolation by quantifying individuals' interaction frequency with family and friends. The LSNS-6 consists of 6 items divided into two dimensions: the first 3 items assessed social relationships with family members (e.g., "How many family members do you meet with at least once a month?"), while the latter 3 items focused on relationships with friends. Responses were scored on a 6-point reverse scale (0-5), with total scores ranging from 0 to 30. Higher scores indicate more severe social isolation.

Subjective Cognitive Decline: The Subjective Cognitive Decline Questionnaire-9 (SCD-9; Jessen et al., 2014) was employed to measure self-perceived cognitive decline. This 9-item scale comprehensively evaluated memory function. Items were scored dichotomously or trichotomously: "Yes" or "Frequently" = 1 point; "Occasionally" = 0.5 points; "No" or "Never" = 0 points. Higher total scores reflected more pronounced subjective cognitive decline symptoms, greater concern about cognitive deterioration.

Depression: depression were assessed using the Patient Health Questionnaire-9 (PHQ-9; Kroenke et al., 2001), a widely validated screening tool for depression. The 9 items correspond to the DSM diagnostic criteria for depression, covering dimensions such as depressed mood, anhedonia, sleep disturbances, fatigue, appetite changes, low self-esteem, concentration difficulties, psychomotor retardation/agitation, and suicidal ideation. Items were rated on a 4-point Likert scale: 0 = "Not at all"; 1 = "Several days"; 2 = "More than half the days"; 3 = "Nearly every day." Total scores range from 0-27, with clinical cut-offs: 5-9 = mild; 10-14 = moderate; 15-19 = moderately severe; ≥20 = severe depression. Higher scores indicated greater depression severity.

For longitudinal data collection, considering older participants' cognitive load during survey completion, we adopted a simplified single-item approach for each variable at each timepoint (e.g., depression was assessed with: "During the lockdown period, how many days in the past two weeks did you experience depressive feelings?"). Furthermore, building on the compensatory hypothesis that digital technology may enhance well-being by facilitating social connections (Kushlev & Leitao, 2020), particularly when face-to-face interactions were abruptly reduced during lockdown. We operationalized social isolation as a multidimensional construct, differentiated social isolation measurements into online versus offline modalities.

**Procedures**

We screened participants who met the inclusion criteria through nationwide online recruitment channels, and after obtaining informed consent, participants completed a comprehensive online questionnaire containing demographics, the Lubben Social Network Scale (LSNS-6), the Subjective Cognitive Disorder Questionnaire (SCD-9), and the Patient Health Questionnaire (PHQ-9), as well as other covariates (e.g., quality of sleep, history of chronic illness, etc.). In addition, in order to assess the relationship of the three factors over time, subjects were asked to report their self-evaluation on social isolation (online and offline), depression, and subjective cognitive decline at three times: during the lockdown period, just after it was lifted, and six months after it was lifted. The study implemented strict quality control measures, including the use of IP address restrictions to prevent duplicate responses, exclusion of responses with unreliable answers (e.g., sleep duration < 0), manual review of questionnaires with short response times (single-question items < 3 seconds), and telephone callbacks to supplement missing data after each survey. Participants who completed all surveys were offered modest compensation and optional mental health referrals were offered to participants screened for major depression (PHQ-9 ≥ 15 points).

**Data analysis**

First, Harman's single-factor test was performed to check for common method bias among the questionnaire items. The results of exploratory factor analysis showed that the amount of variance explained by the first factor was 26.75% (below the critical value of 30%), a result that confirms that the data in this study do not suffer from a serious problem of common method bias (Malhotra et al., 2006). Second, we calculated correlation coefficients between social isolation, depression, and subjective cognitive decline. Third, we used a network analysis to examine the relationships between social isolation, subjective cognitive decline, and depression across items. This analysis aimed to determine which items played a significant role in the relationship between these three variables and their clustering. The network analysis used the six items of social isolation, the nine items of subjective cognitive decline, and the nine items of depression as nodes in the network. The structure of the network was visualized using the Fruchterman-Reingold algorithm, which places the more influential nodes closer to the center of the network, thereby determining which variable plays a central metric role in the network. In order to comprehensively assess the network characteristics, we focus on four key centrality metrics: Betweenness centrality reflects the ability of a node to act as a "hub" connecting different network regions; Strength centrality characterizes the overall degree of a node's connection to other nodes; Closeness centrality measures the efficiency of information dissemination of a node; Expected influence evaluates the potential influence of a node on the whole network system. Further, we also computed the bridging metrics of the network. All network analyses were done using the bootnet and qgraph packages in the R language environment, and the stability of the network structure and the reliability of the centrality metrics were ensured by 1000 bootstrap sampling tests. In addition, node centrality discrepancy tests were performed using botnet, an R package, to check for differences in node strength. Node strength values and reliability of connections were determined by correlation stability coefficients. Values greater than 0.25 indicate moderate stability and values greater than 0.5 indicate strong stability. This method of analysis not only identifies direct associations between scale items, but also reveals potential core symptoms and key items. It provides a novel network perspective for understanding the mechanisms of interaction between social isolation, cognitive functioning, and depressed mood.

    Finally, in order to deeply explore the dynamic causal relationship between social isolation, depression and subjective cognitive decline, the present study used the Cross-Lagged Mediation Model (CLMM) to analyze the longitudinal interaction mechanisms among the three. Specifically, we constructed a structural equation model containing three time points (T1-T3), focusing on the mediating role of depression in the relationship between SI and SCD. During data analysis, all data entry and preliminary processing was done using SPSS 26.0 (IBM Corporation, Armonk, NY, USA), whereas more complex cross-lagged mediation analyses were realized with Mplus 7.3 (Muthén & Muthén, 1998-2017) software. To deal with the inevitable

problem of missing data in the study, we used maximum likelihood estimation in conjunction with robust standard errors for model estimation, which is able to maintain the validity of parameter estimates even when the data are not normally distributed or when there are missing values. In terms of results presentation, all path coefficients are standardized to facilitate effect size comparisons, with reference to the criteria (Orth et al., 2022): 0.03 for small effect sizes, 0.07 for medium effect sizes, and 0.12 for large effect sizes. This analytic strategy not only controls for autoregressive effects of the variables, but also simultaneously assesses the predictive effect of SI on depression (T1→T2) and the effect of depression on subsequent SCD (T2→T3), thus providing strong statistical evidence for understanding the causal temporal relationship of the three. In addition, model fit was assessed by a combination of several metrics, including the comparative fit index (CFI), Tucker-Lewis index (TLI), root mean square error of approximation (RMSEA), and standardized root mean square residual (SRMR), to ensure that the theoretical model was in good agreement with the observed data.

**Results**

First, Pearson correlation analysis revealed significant positive correlations between social isolation and subjective cognitive decline ($r = 0.221$, $p < 0.01$), between social isolation and depression ($r = 0.247$, $p < 0.01$), and between subjective cognitive decline and depression ($r = 0.459$, $p < 0.01$). Network analysis results demonstrated complex interaction patterns among them (Figure 1). Visualization of the network structure showed predominantly positive correlations (green connecting lines) between all variable nodes, with the thickness of the connecting lines intuitively reflecting the strength of associations between variables. Notably, item 9 of the depression scale (PHQ9, concerning "suicidal or self-harm ideation") exhibited prominent centrality characteristics, with its node position located in the core region of the entire network. Quantitative analysis of centrality metrics (Figure 2) further confirmed that PHQ9 ranked first across three dimensions: (betweenness centrality = 3.79; closeness centrality = 2.55; connection strength = 2.36), indicating that this node served both as the core driver of depressive symptom clusters and as a crucial hub for cross-community (SI - SCD) information exchange. This further verified PHQ9 as an important connecting variable in the network, providing evidence for considering depression as a potential psychological mechanism linking social isolation and subjective cognitive decline.

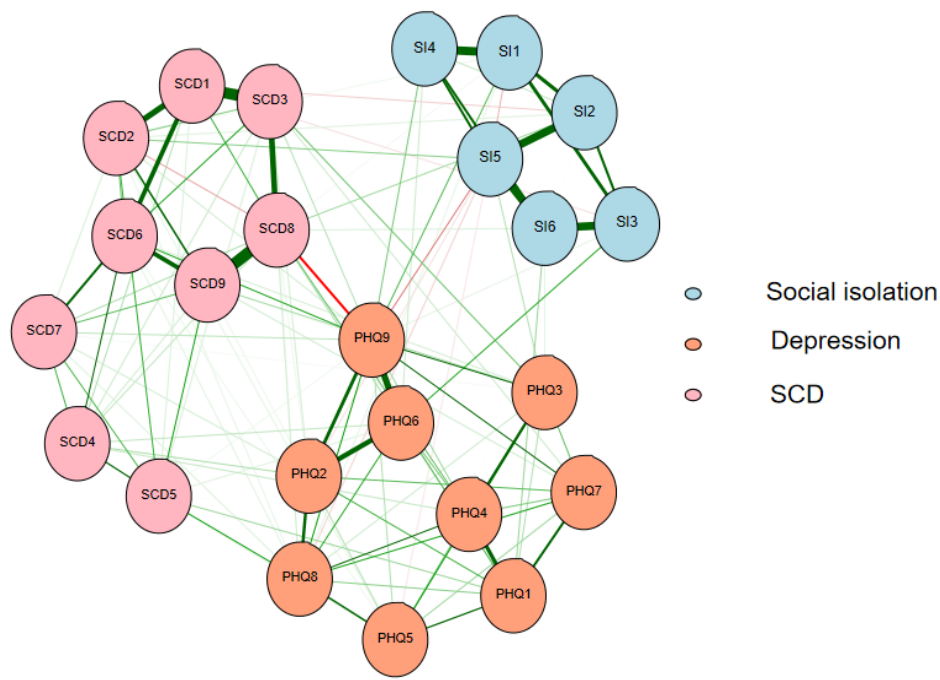

**Figure 1**. Network diagram of item-level associations among social isolation (SI),

depression (PHQ), and subjective cognitive decline (SCD) in older adults. *Note*: In the figure, pink nodes represented subjective cognitive decline, orange nodes represented depression, and light blue nodes represented social isolation. Nodes with stronger correlations were positioned closer to each other. The thickness of the edges indicated the strength of the correlations.

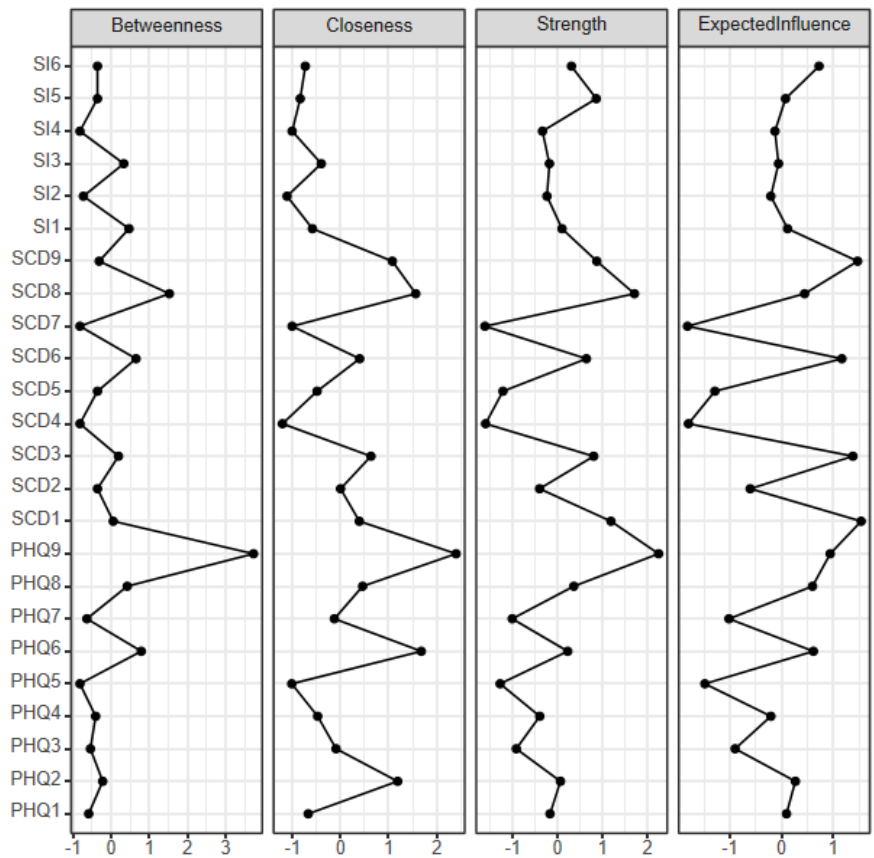

**Figure 2**. The results of centrality metrics analysis for the network.

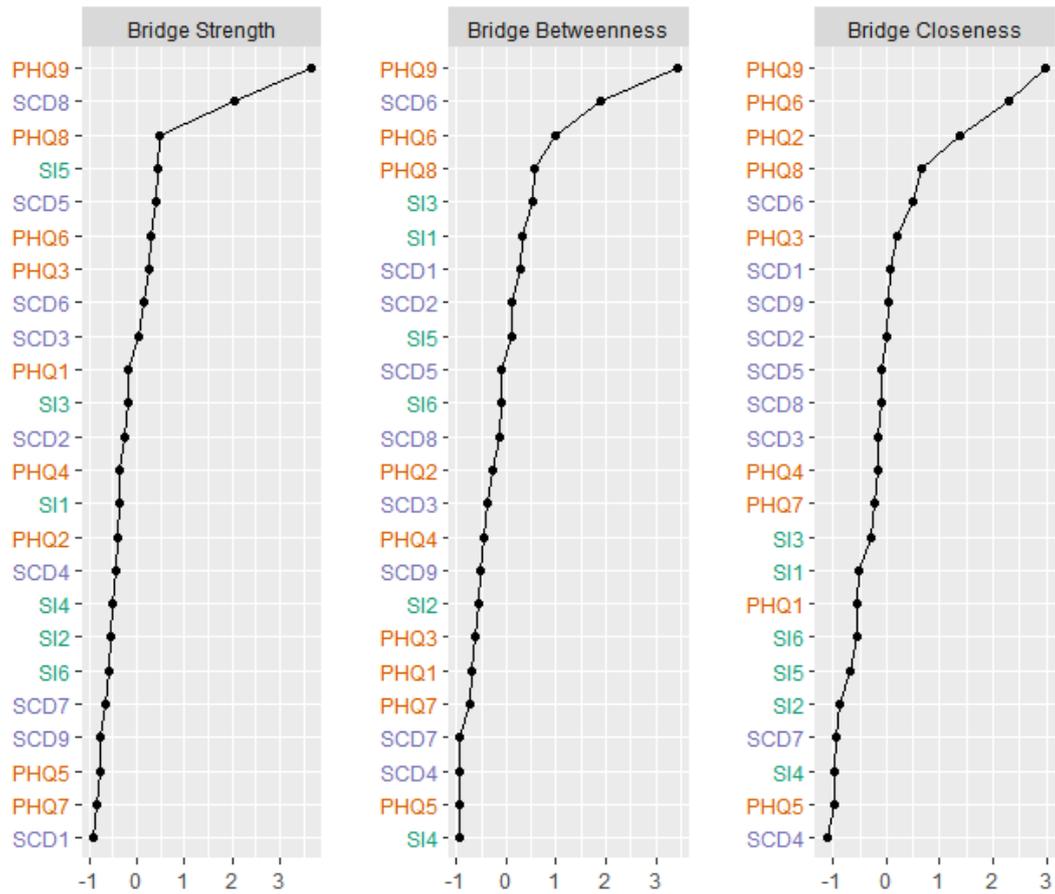

**Figure 3**. The results of bridge centrality analysis for the network.

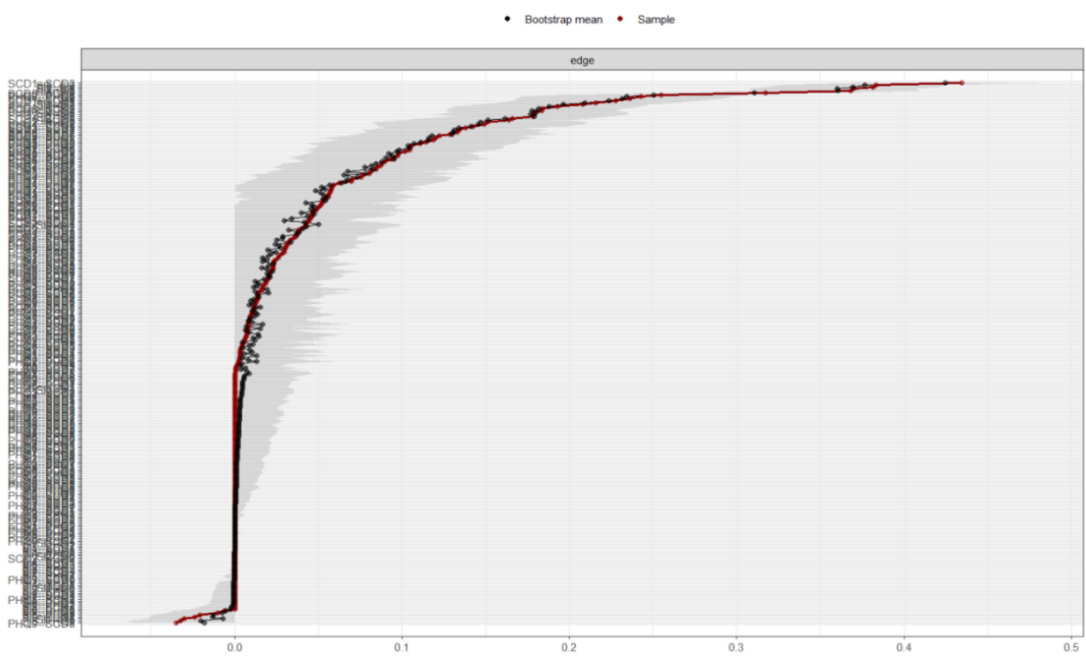

**Figure 4**. Estimated edge weights with 95% confidence intervals (CIs) for the network.

Bridge centrality analysis (Figure 3) provided further interpretation of the network structure. PHQ9 demonstrated the highest bridge strength (bridge strength = 3.72) among all nodes, ranking first, suggesting its potential role as a mediating node connecting the social isolation cluster (composed of LSNS-6 items) with the cognitive impairment cluster (composed of SCD items). This finding offered network-level evidence supporting the "depression mediation hypothesis" - that social isolation may lead to subjective cognitive decline by exacerbating depression.

Figure 4 displayed the network edge weighted and their 95% confidence intervals estimated via bootstrap methods. Edge weight values reflected the strength of associations between variables, while confidence intervals assessed the statistical significance of these associations. The results showed that at the 95% confidence level, the majority of confidence intervals for variable associations did not include zero, indicating significant differences between edge weight estimates and zero ($p < 0.05$). These significant edge associations suggested that there existed real and stable interrelationships among measurement items in the network, with different dimensions potentially influencing and maintaining each other through specific interaction pathways.

Finally, we constructed a cross-lagged mediation model (Figure 5) using a three-wave longitudinal design to systematically examine the mediating role of depression in the relationship between social isolation(both online and offline) and subjective cognitive decline. As shown in Table 2, descriptive statistics across the three time points revealed dynamic trends in the variables. Path analysis results in Figure 5(a) demonstrated that: 1)T1 online social isolation significantly predicted T2 depression ($\beta = -0.02$, $p = 0.009$); 2) T2 depression significantly predicted T3 subjective cognitive decline ($\beta = 0.09$, $p = 0.023$); Bootstrap testing (1000 samples) confirmed a significant longitudinal mediating effect of depression (indirect effect = -0.002, $p = 0.042$). The model showed good fit indices ($\chi^2 = 162.470$, $df = 22$, $RMSEA = 0.072$, 90% $CI = [0.062, 0.083]$, $CFI = 0.973$, $TLI = 0.960$, $SRMR = 0.041$). Path analysis results in Figure 5(b) show that: T1 offline social isolation significantly predicted T2 depression ($\beta = -0.02$, $p = 0.009$); T2 depression significantly predicted T3 subjective cognitive decline ($\beta = 0.09$, $p = 0.024$). This model also demonstrated good fit indices ($\chi^2 = 152.383$, $df = 20$, $RMSEA = 0.073$, 90% $CI = [0.063, 0.084]$, $CFI = 0.975$, $TLI = 0.959$, $SRMR = 0.036$). These results supported the potential pathway of "reduced social support (increased social isolation) → worsened depression → increased subjective cognitive decline". Consistent findings for both online and offline social isolation suggested that the impact of social isolation maintains cross-context stability. Whether through lack of online social interaction or reduced real-world offline social contact, both leaded to subjective cognitive decline by exacerbating depression, indicating that the negative mental health effects of social isolation were universal and not limited by the form of social interaction (virtual/in-person). Furthermore, these findings validated the crucial mediating role of depression in mental health issues among older adults, providing dynamic supportive evidence for this pathway.

**Table 2**. Descriptive statistics of social isolation, depression, and subjective cognitive decline in the longitudinal study (*N* = 1230)

|  | *Mean* (*SD*) |
|---|---|
| **During Lockdown Period (T1):** | |
| Online social isolation | 3.44(1.14) |
| Offline social isolation | 2.87(1.12) |
| Subjective cognitive decline | 2.73(0.63) |
| Depression | 0.56(0.64) |
| **Immediately After Lockdown Lifting (T2):** | |
| Online social isolation | 3.59(1.18) |
| Offline social isolation | 3.32(1.12) |
| Subjective cognitive decline | 2.46(0.76) |
| Depression | 0.42(0.57) |
| **6 Months Post-Lockdown (T3):** | |
| Online social isolation | 3.66(1.21) |
| Offline social isolation | 3.44(1.17) |
| Subjective cognitive decline | 2.32(0.81) |
| Depression | 0.39(0.57) |

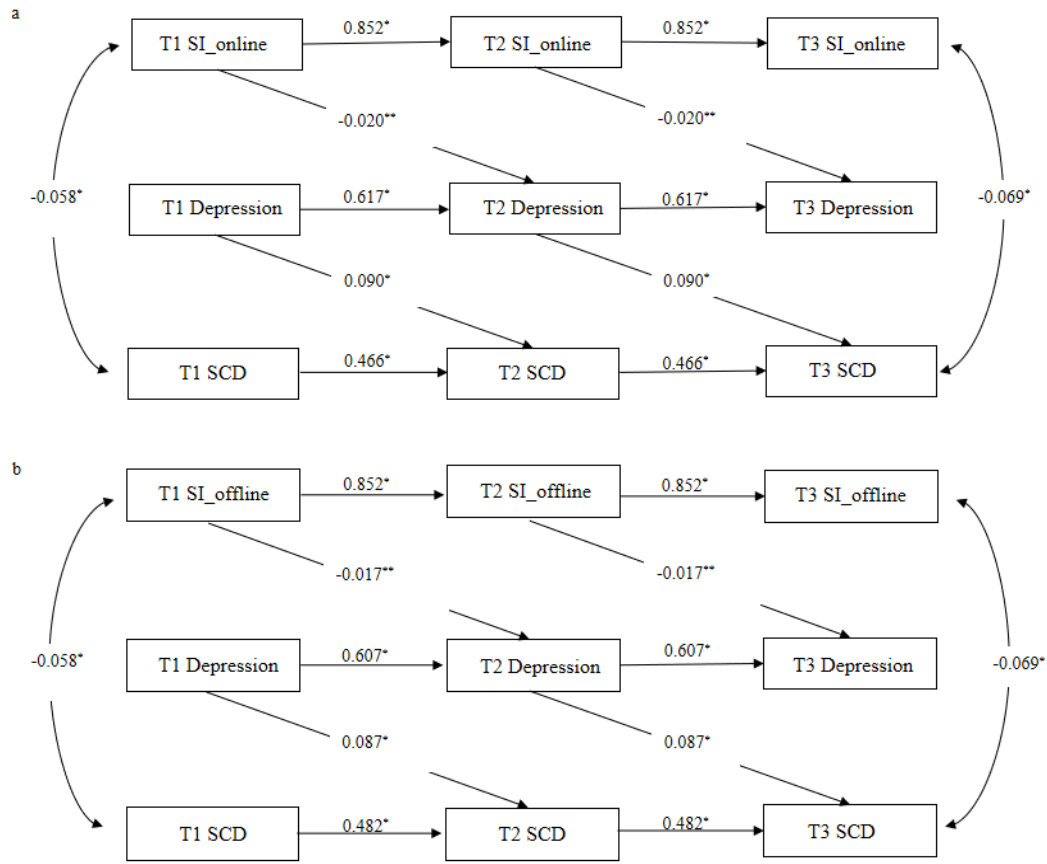

**Figure 5**. Cross-lagged mediation model (CLMM) after controlling for age, gender, and education. a) CLMM for the relationship between online social isolation, depression, and subjective cognitive decline; b) CLMM for the relationship between offline social isolation, depression, and subjective cognitive decline; *Note*: *$p < 0.05$, **$p < 0.01$; Showing all significant standardized path coefficients).

**Discussion**

This study systematically revealed the interaction mechanisms among social isolation, depression, and subjective cognitive disorder by integrating network analysis and longitudinal modeling methods. The results of the network analysis showed that depression (especially the suicidal ideation items in the PHQ-9 scale) exhibited significant centrality characteristics throughout the psychological symptom network, and its multiple bridging centrality indicators were significantly higher than those of the other nodes, confirming the potential mediating role of depression between social isolation and subjective cognitive disorder. The longitudinal cross-lagged mediation model further showed that at the three time points of during sequestration, just released from sequestration, and 6 months after release from sequestration, social isolation levels at the former time point significantly predicted subsequent depression, which in turn predicted later subjective cognitive decline. This mediation pattern was consistently observed for both online and offline social isolation, with good model fit indices. These findings supported the longitudinal effect model of "increased social isolation → increased depression → increased subjective cognitive decline" and provide important evidence for understanding the psychosocial mechanisms of cognitive decline in older adults.

  The present study found that depression (particularly suicidal ideation) exhibited significant centrality and bridging roles in the psychological network through network analysis methods, consisting with previous studies and further confirming the key pivotal role of depression between social isolation and cognitive decline. Specifically, older adults with social support deprivation tend to exhibit more severe depression (Cornwell & Laumann., 2015; Tian & Li., 2023), which, in turn, lead to more negative evaluations of their cognitive abilities. Notably, suicidal ideation, a core symptom of depression, exhibits the highest intensity centrality across the entire network of psychological symptoms. According to Joiner's interpersonal theory (Chu et al., 2017), suicidal ideation is closely related to frustrated sense of belonging and perceived tiredness. In the context of a major public health event such as the COVID-19 outbreak, older adults may be more susceptible to perceptions of being a burden to family and society, which may exacerbate suicidal ideation and other successive psychological reactions.

The longitudinal analyses robustly confirmed the stable mediation pathway (social isolation → depression → cognitive decline) across all phases and interaction modalities, demonstrating both immediate and lasting impacts of social disconnection on older adults' cognitive health. This important finding not only supports the "multidimensional theory of social connectedness" that different forms of social isolation may negatively affect mental health(Qualter et al., 2015), but also confirms the cross-situ robustness of the effects of social isolation on cognitive functioning from a longitudinal perspective. Older adults with reduced social connections are at higher risk for depression, and that such effects are not only immediate, but may also have far-reaching long-term consequences.

**Conclusion**

The present study demonstrates that social isolation due to lockdown exacerbates depression (especially suicidal ideation) and thus subjective cognitive decline in older adults. depression consistently played a key mediating role between social isolation and cognitive problems, and this role was not affected by online or offline forms of socialization. This suggests that in order to promote older adult mental health and increase older adults' self-confidence in their own cognitive appraisal, a shift should be made from the traditional single perspective to a multidimensional model that integrates social support reinforcement, depressive symptom alleviation, and cognitive functioning preservation to produce a broader network-improvement effect, leading to more effective health promotion.